# Superlinear growth of Rayleigh scattering-induced intensity noise in single-mode fibers

James P. Cahill,[1,2,*] Olukayode Okusaga,[1] Weimin Zhou,[1] Curtis R. Menyuk,[2] and Gary M. Carter[2]

[1]*U.S. Army Research Laboratory, 2800 Powder Mill Rd. Adelphi, MD 20783, USA*
[2]*Dept. of Comp. Sci. and Elec. Eng., University of Maryland: Baltimore County, Baltimore, MD 21250, USA*
**james.p.cahill.ctr@us.army.mil*

**Abstract:** Rayleigh scattering generates intensity noise close to an optical carrier that propagates in a single-mode optical fiber. This noise degrades the performance of optoelectronic oscillators and RF-photonic links. When using a broad linewidth laser, we previously found that the intensity noise power scales linearly with optical power and fiber length, which is consistent with guided entropy mode Rayleigh scattering (GEMRS), a third order nonlinear scattering process, in the spontaneous limit. In this work, we show that this behavior changes significantly with the use of a narrow linewidth laser. Using a narrow linewidth laser, we measured the bandwidth of the intensity noise plateau to be 10 kHz. We found that the scattered noise power scales superlinearly with fiber length up to lengths of 10 km in the frequency range of 500 Hz to 10 kHz, while it scales linearly in the frequency range of 10 Hz to 100 Hz. These results suggest that the Rayleigh-scattering-induced intensity noise cannot be explained by third-order nonlinear scattering in the spontaneous limit, as previously hypothesized.

**1. Introduction**

When light travels in a material, inhomogeneities in the material cause the light to scatter. We can represent the inhomogeneities that scatter light with the permittivity. Fluctuations that occur on a molecular scale alter the tensor component of the permittivity, leading to Raman and Rayleigh-wing scattering [1]. Fluctuations of the thermodynamic properties—i.e. on the scale of many molecules—affect the scalar component of the permittivity. In this case, fluctuations in the density are the primary source of fluctuations in the permittivity, and other thermodynamic properties scatter light indirectly, via changes in the density [2]. For example, isentropic density fluctuations induced by pressure fluctuations lead to Brillouin scattering, and isobaric density fluctuations induced by entropy fluctuations lead to Rayleigh scattering [3]. In this paper, we will focus on Rayleigh scattering.

By the laws of thermodynamics, entropy-induced density fluctuations naturally exist in a material at thermodynamic equilibrium. The scattering caused by these fluctuations is called spontaneous Rayleigh scattering [1]. In optical fibers and other glasses, there is a second source of entropy-induced density fluctuations [4]. When the glass is formed at high temperature, entropy-induced density fluctuations occur with a higher variance than they do at room temperature. When the glass is cooled, the high-temperature fluctuations are frozen into the structure of the glass and lead to a greater power of scattered light than those induced at room temperature. Scattering from frozen-in fluctuations is often cited as the dominant loss mechanism in optical fibers [5].

Entropy-induced density fluctuations in a material may also be induced by light that passes through it. This effect is called stimulated Rayleigh scattering (STRS). In fluids and gases, feedback between the density fluctuation and incident and scattered light causes exponential growth of the scattered power with increasing incident power and interaction length [6,7]. The bandwidth of STRS is determined by the time for an entropy fluctuation to dissipate. In STRS, the material fluctuation is diffusive, rather than propagating; so, in contrast to other stimulated processes, there is no frequency offset due to wave emission. Nonetheless the scattered power does have an offset that reaches a maximum at half its bandwidth when measured with a narrow linewidth laser and at half the laser linewidth when measured with a broad linewidth laser [6,7].

In optical fibers, heat diffusion occurs across the transverse plane of the fiber core, leading to a bandwidth, and hence apparent frequency shift, that is approximately 1000 times smaller than the value predicted in bulk silica [8,9]. Because of the influence of the waveguide on the bandwidth, this type of scattering is called guided entropy mode Rayleigh scattering (GEMRS) [8].

Light that scatters from the frozen-in Rayleigh fluctuations can also cause optical intensity noise via conversion of the laser phase noise [10,11]. In the limit that the phase noise of the light-source is large relative to the delay induced by the optical fiber, the intensity noise spectrum will have the form of a Lorentzian, with a bandwidth corresponding to the linewidth of the light source. To the best of our knowledge, the opposite limit has not been analyzed in detail.

We previously measured the intensity noise spectrum of the light scattered backwards in an optical fiber using a low-RIN distributed feedback (DFB) laser with a linewidth of 400 kHz [8,12]. We measured a plateau of increased noise power extending from 10 Hz to 100 kHz. We have shown that this noise can degrade the performance of optoelectronic oscillators (OEOs) [13], broadband analog RF-photonic fiber links [14], and narrowband analog RF-photonic fiber links that are used for frequency transfer [15]. Other groups have measured Rayleigh-scattering-induced noise at similar offset frequencies and shown that it limits the performance of fiber-based gyroscopes [16,17]. Rayleigh scattering also causes modal instability in fiber amplifiers and lasers [18].

We determined that the power in this plateau scales linearly with the input optical power and the optical fiber length. We also found using heterodyne measurements that the power spectral densities for positive-frequency and negative-frequency scattering were equal. Based on these results, we suggested that the noise could be explained by third order nonlinear scattering in the spontaneous limit and concluded that the scattering mechanism with the appropriate bandwidth is GEMRS [12].

In this paper, we will demonstrate that the laser phase noise is a significant factor in determining the behavior of the Rayleigh-scattering-induced intensity noise. When we measure the spectrum with a laser that has phase noise $10^4$ times lower than the DFB laser used in [12], we find that the intensity noise plateau extends only to 10 kHz, a factor of 10 narrower than when measured with the standard DFB laser. Additionally, when using the low phase noise laser, the intensity noise power scales superlinearly with fiber length up to 10 km over the frequency range 500 Hz – 10 kHz, although it scales linearly in the frequency range 10 Hz – 100 Hz. We also find that the intensity noise power scales linearly with the input power in all frequency ranges. The results presented in this paper are inconsistent with the conclusions of [12], indicating that a new physical picture is required to explain all of the experimental results.

## 2. Experimental Setup

The experimental apparatus, shown in Fig. 1, is similar to the one used in [8].

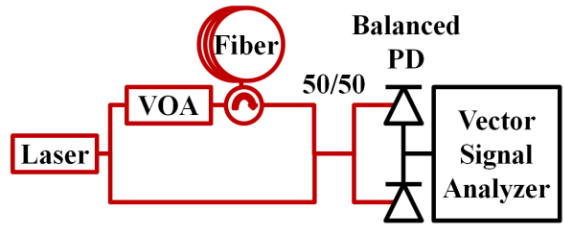

Fig. 1 Experimental apparatus used to measure the intensity noise of the backward-scattered light.

The laser under test generates the light for the system, and a 75%-25% coupler splits the light into two arms. One arm (the bottom arm in Fig. 1) serves as a local oscillator, while the other (the top arm in Fig. 1) guides the light through a variable optical attenuator (VOA) and into an optical circulator. The circulator directs the light to the fiber under test (FUT), which is enclosed in a box that is designed to isolate the fiber from environmental fluctuations. Scattering occurs in the FUT, and the light that scatters backwards travels back through the circulator and into a two-by-two 50% coupler. The backward scattered light combines with the local oscillator in the coupler, and the two outputs illuminate the diodes of a 50-Ω terminated balanced photodetector (Balanced PD). The vector signal analyzer digitizes the voltage across the termination resistor and computes the power spectral density. We have calibrated this power spectral output to find the absolute power spectral density of the scattered light in dBm/Hz.

We measured the intensity noise of the backward scattered light using three different lasers. The first laser, denoted "NLW-FL," is an NKT Photonics Koheras Adjustik fiber laser that has a linewidth of less than 100 Hz. The second laser, denoted "NLW-DFB," is a Teraxion NLL. It is a narrow linewidth, grating-stabilized DFB laser that has a linewidth of less than 5 kHz. The third laser, denoted "DFB," is an EM4 high power DFB laser, the same model that was used in [8]. It has a linewidth of approximately 400 kHz. We note that although the linewidths of the narrow linewidth lasers, NLW-FL and NLW-DFB, are specified to be more than an order of magnitude apart, the power spectral density of the phase noise of these two lasers is specified to be approximately the same in the present region of interest—i.e., offset frequencies 10 Hz – 1 MHz. As a consequence, we do not expect the NLW-FL and NLW-DFB to behave differently due to their different linewidths. Instead, we expect any differentiation in their spectra to result from the laser relative intensity noise. By contrast, in the case of the broad linewidth laser, DFB, the power spectral density of the phase noise is $10^4$ times higher than that of the narrow linewidth lasers between offset frequencies 10 Hz – 1 MHz. Hence, we expect any effect of the laser phase noise to differentiate this spectrum from the spectra that result from the narrow linewidth lasers.

### 3. Results

First, we measured the length dependence of the intensity noise spectrum of the backward scattered light using each laser. Figure 2(a) shows the spectrum created by the NLW-FL at different lengths of fiber. The noise floor of the measurement system (black curve) was measured with no optical fiber added to the optical circulator. At offset frequencies 10 Hz – 1 kHz, the noise floor was determined by the vector signal analyzer, and at offset frequencies of 1 kHz – 1 MHz, it was determined by the laser RIN. The measured intensity noise spectra were limited by the noise floor for offset frequencies of 100 kHz – 1 MHz.

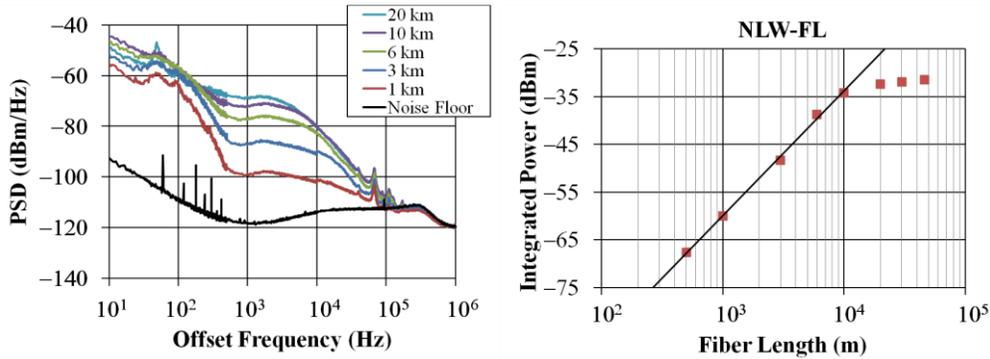

Fig. 2. (a) Intensity noise spectra using NLW-FL for several fiber lengths. (b) Integrated noise power in the frequency range 500 Hz to 10 kHz as a function of fiber length. The data is taken with the NLW-FL. The slope of the linear fit (black line) is 2.6.

The intensity noise spectra can each be divided into two distinct frequency regions, according to the behavior of the scattering. The power in each region scales differently with increasing fiber length. The first region comprises offset frequencies from 10 Hz to 100 Hz. In this region, a distinct plateau exists at all fiber lengths. To quantify how the noise power in this region scales with fiber length, we integrated the scattered power in the spectral region from 10 Hz to 100 Hz. We found that the integrated power scales linearly with the fiber length.

The second region comprises offset frequencies from 500 Hz to 10 kHz. In this region, a distinct plateau grows as the fiber length increases. We integrated the scattered power in the spectral region from 500 Hz to 10 kHz. The results for the NLW-FL are shown in Fig. 2(b). With logarithmic axes, the integrated power appears linear, with a slope of 2.6, for fiber lengths from 500 m to 10 km. For optical fiber lengths of 20 km and above, the integrated power increases with a slope of 0.2. We note that the technique of finding the slope of the line on a logarithmic plot is not a precise method to determine the functional dependence of one parameter on another, since many different functions may have the same slope over certain domains. However, this method does provide a lower bound on the degree of the polynomial that best describes the data. Hence, this result indicates that the integrated power is related superlinearly, but not exponentially to the fiber length.

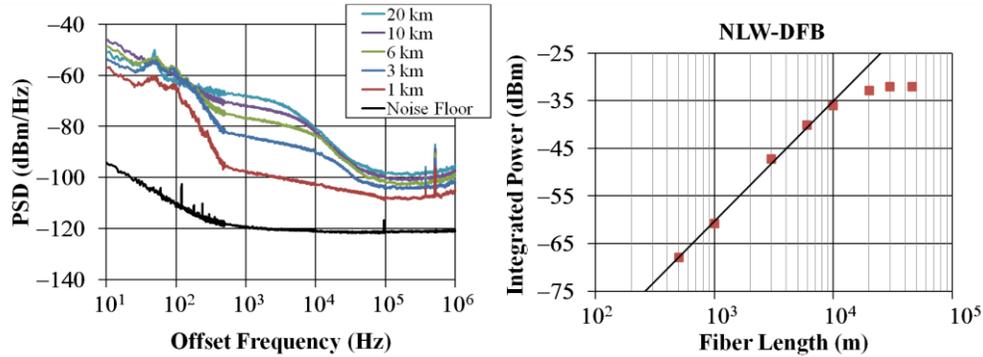

Fig. 3. (a) Intensity noise spectra using NLW-DFB for several fiber lengths. (b) Integrated noise power in the frequency range 500 Hz to 10 kHz as a function of fiber length. The data is taken with the NLW-DFB. The slope of the linear fit (black line) is 2.5.

Figure 3(a) shows the intensity noise spectrum created by the NLW-DFB laser in various lengths of fiber. The noise floor of the measurement system (black curve) was measured with no optical fiber added to the optical circulator. The noise floor was determined by the vector signal analyzer noise at all offset frequencies. The measured intensity noise spectra were not limited by the noise floor.

The intensity noise spectra can each be divided into two distinct frequency regions, similar to the NLW-FL spectra. The integrated power from 10 Hz – 100 Hz increases linearly with fiber length. The integrated power in the second plateau for the NLW-DFB laser is shown in Fig. 3(b). The slope of the linear fit is 2.5. Hence, as with the NLW-FL, the power increases superlinearly with fiber length, up to 10 km. For optical fiber lengths 20 km and above, the power increases more slowly with the fiber length, with a slope of 0.2. We find that the qualitative behavior of the NLW-DFB-induced spectrum is the same as the behavior of the NLW-FL-induced spectrum. This result indicates that the laser RIN does not affect the scattering spectrum for offset frequencies between 10 Hz and 100 kHz.

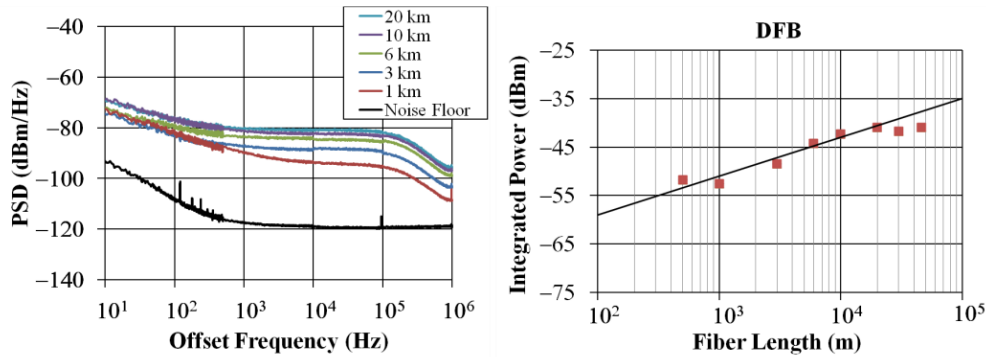

Fig. 4. (a) Intensity noise spectra using the broad linewidth DFB laser for several fiber lengths. (b) Integrated noise power in the frequency range 500 Hz to 10 kHz as a function of fiber length. The data is taken with the broad linewidth DFB laser. The slope of the linear fit (black line) is 0.8.

Figure 4(a) shows the intensity noise spectrum created by the broad linewidth DFB laser for various lengths of fiber. The noise floor of the measurement system (black curve) was measured with no optical fiber added to the optical circulator. Similar to the case of the NLW-DFB laser, the noise floor was determined by the vector signal analyzer noise at all offset frequencies. The measured intensity noise spectra were not limited by the noise floor.

For the high-phase-noise DFB laser, the spectrum can no longer be divided into distinct regions. The spectra are similarly shaped at all offset frequencies for the fiber lengths measured. For comparison to the narrow linewidth lasers, we have again integrated the scattered power from 500 Hz to 10 kHz, as shown in Fig. 4(b). In this case, the scattered power scales linearly with fiber length, as we found in [12].

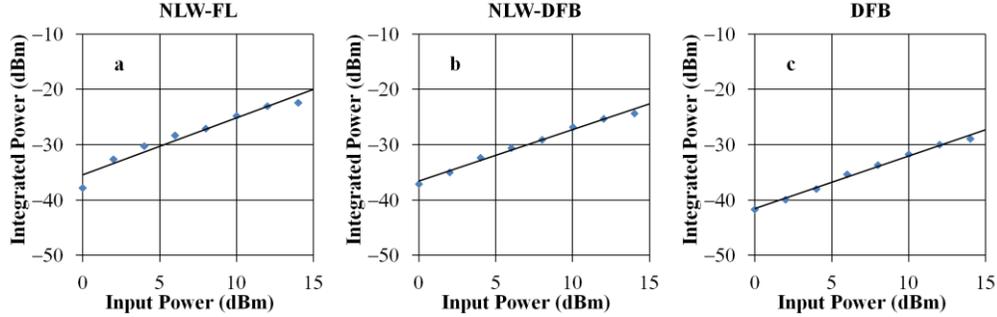

Fig. 5. Integrated noise power from 500 Hz to 10 kHz as a function of input power for each laser. The slopes of the linear fits (black lines) are 1.0 (a), 0.9 (b), and 0.9 (c).

Next, we measured the scaling of the intensity noise as a function of the input optical power, while holding the fiber length constant at 10 km. The shape of the scattered spectrum does not change with the input power for any laser. The integrated power in the frequency range 500 Hz to 10 kHz for the NLW-FL, NLW-DFB, and broad linewidth DFB lasers are shown in Figs. 5(a) – (c), respectively. The scattered power increases linearly with the input optical power for all three lasers. This behavior remains the same for the optical fiber lengths measured—i.e. 500 m – 10 km.

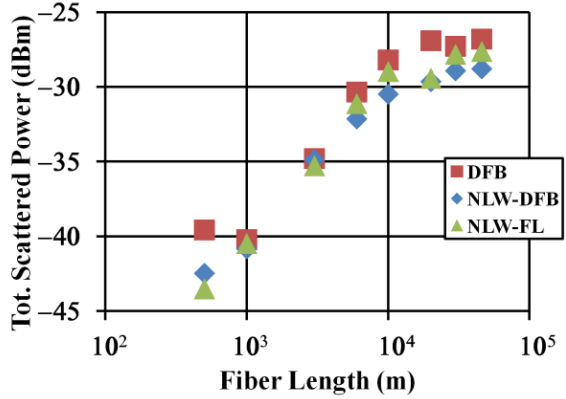

Fig. 6. Total integrated power scattered backwards for each laser, with 0 dBm input power.

We note that despite the differences in scaling with fiber length, the total power scattered by each laser is approximately the same. Figure 6 shows the total backward scattered power with different fiber lengths for each laser. All data in this figure was taken with an input power of 0 dBm. The total power that scatters from each laser is within 3 dB of the other lasers for all fiber lengths. For the NLW-FL and NLW-DFB lasers, the integrated power is greater in the first plateau than in the second plateau for all fiber lengths, so that the superlinear growth with fiber length does not significantly alter the total integrated power.

Figure 7 directly compares the intensity noise spectra of the backward scattered light from each laser. The data was measured in 10 km single mode fiber with an input power level of 0 dBm. The spectra obtained with the two narrow linewidth lasers are nearly identical for offset frequencies less than 100 kHz. By contrast, the broad linewidth DFB-induced scattering exhibits lower power spectral density than the other lasers at offset frequencies less than 10 kHz and higher power spectral density at offset frequencies greater than 10 kHz.

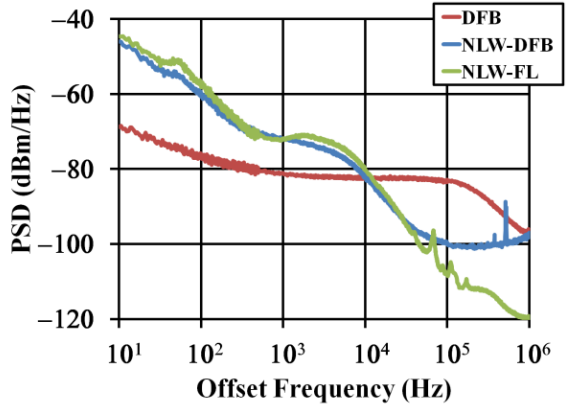

Fig. 7. Intensity noise spectra for each laser in 10 km of fiber with 0 dBm input power.

By comparing the scattered spectra measured with each laser, we can see that the additional phase noise of the broad linewidth DFB laser has spread the noise power over a wider bandwidth.

## 4. Discussion of Experimental Results

We can combine the experimental results reported in this paper with the results found in [8,12] in order to summarize the most notable characteristics of the scattering process:
1. The bandwidth of the intensity noise is between 10 kHz and 100 kHz, and its value depends on the laser phase noise.
2. The intensity noise spectrum is symmetric about the optical carrier—i.e., it has the same power spectrum at a given positive offset frequency as it does at the corresponding negative offset frequency.
3. The intensity noise power grows in direct proportion to the incident power.
4. The intensity noise power grows superlinearly, but not exponentially, with the fiber length in the offset frequency region 500 Hz – 10 kHz for lasers with low phase noise (see Figs. 2 and 3) and grows in direct proportion to the fiber length for lasers with higher phase noise (see Fig. 4).
5. The scaling of the intensity noise power with optical fiber length changes when the optical fiber length exceeds 10 km.
6. The laser intensity noise has little effect on the scattered spectrum in the range of 10 Hz – 100 kHz.

These trends rule out some potential theoretical explanations. In [8], we concluded that the backward-scattered intensity noise is caused by a third order light scattering effect, in which the laser intensity noise experiences frequency-dependent gain. This picture corresponds to the well-known result, as reported in, e.g. [1, p. 440, eq. 9.3.25]:

$$I_r(0) = I_r(L)\exp\left[gI_i L\right]. \tag{1}$$

Where $I_i$ is the intensity of the pump laser, $I_r(L)$ is the seed intensity at the end of the fiber, $g$ is the real-valued frequency-dependent gain factor, and $L$ is the fiber length. We note that this result is valid only under the assumption that there is no pump depletion. This assumption is valid in the case that the loss of the optical fiber is well-described by the propagation loss factor of 0.2 dB/km, which is the case for all of the experimental results reported in this paper and in [8,12]. Based on this notional model, we expect the scattered power to grow in direct proportion to the incident power and fiber length when the product $gI_iL$ is much less than one—i.e, in the spontaneous limit—and exponentially with the incident power and fiber length when the product $gI_iL$ is above a threshold level—i.e., the stimulated limit. When we use a waveguide-narrowed Rayleigh gain factor, the spontaneous limit of this notional model matches experimental observations 1 through 3, as shown in [8,12].

However, the growth patterns with length that are reported in this paper are not consistent with either the spontaneous or the stimulated limit of third order nonlinear scattering, and thus are inconsistent with the conclusions of [8,12]. Instead, superlinear growth within a limited frequency band suggests that a frequency-dependent interaction of the laser phase noise, such as the one described in [10,11]. Further theoretical study, combined with additional experiments, will be required to definitively establish the physical mechanism that causes all of the observed characteristics of the intensity noise.

## 5. Conclusion

In this paper, we have presented evidence that the backward scattered intensity noise at low offset frequencies is not well-described as third-order nonlinear scattering in the spontaneous limit. We have shown that the behavior of this intensity noise depends on the incident laser phase noise. Using low phase noise lasers, we found that the scattered power in the plateau between 500 Hz and 10 kHz grows superlinearly with fiber length up to 10 km and linearly with input optical power. These scaling patterns are inconsistent with interpretation of this noise as either a standard stimulated or spontaneous process, as proposed in [8,12] on the basis of the experimental data that was then available. We have outlined the most notable characteristics shown by the experimental results in this paper, as well as in [8,12]. Rigorous theoretical work, combined with further experiments, is required to make definitive claims about the physical process that underlies this noise source.